# Towards an Understanding of Developers' Perceptions of Transparency in Software Development: A Preliminary Study

Humphrey O. Obie*, Juliet Ukwella†, Kashumi Madampe*, John Grundy*, Mojtaba Shahin‡
*Monash University, Melbourne, Australia
†Durham College, Ontario, Canada
‡RMIT University, Melbourne, Australia
{humphrey.obie, kashumi.madampe, john.grundy}@monash.edu, juliet.ukwella@dcmail.ca, mojtaba.shahin@rmit.edu.au

*Abstract*—Software applications play an increasingly critical role in various aspects of our lives, from communication and entertainment to business and healthcare. As these applications become more pervasive, the importance of considering human values in software development has gained significant attention. In this preliminary study, we investigate developers' perceptions and experiences related to human values, with a focus on the human value of *transparency*. We interviewed five experienced developers and conducted thematic analysis to explore how developers perceive transparency, violations of transparency, and the process of fixing reported violations of transparency. Our findings reveal the significance of transparency as a fundamental value in software development, with developers recognising its importance for *building trust*, *promoting accountability*, and *fostering ethical practices*. Developers recognise the negative consequences of the violation of the human value of transparency and follow a systematic process to fix reported violations. This includes investigation, root cause analysis, corrective action planning, collaborative problem-solving, and testing and verification. These preliminary findings contribute to the understanding of transparency in software development and provide insights for promoting ethical practices.

## I. INTRODUCTION

As software applications become ever more pervasive, the importance of considering human values in software development has gained significant attention [1], [2]. Human values are the guiding principles of what people consider important in life [3]. Human values encompass a broad range of principles, ethics, and moral considerations that guide our interactions, decisions, and behaviours [4]. Incorporating human values into software development ensures that the resulting applications align with ethical standards, promote user trust, and contribute to the well-being of individuals and society [5], [6].

One important aspect of human values in software development that has not been researched very much to date is *transparency*. Transparency is an attribute of communication in software development that enables stakeholders to answer their questions about the software system during its software life cycle [7], [8]. Transparency also encompasses openness, clarity, and visibility of the inner workings, processes, and actions of software applications [9]. Transparent software applications provide users with insight into how their data is collected, used, and protected. They enable users to understand the algorithms and decision-making processes behind automated systems. Transparency empowers users, promotes accountability, and fosters trust between developers, users, and other stakeholders [7]–[9].

Understanding how developers perceive, address, and prioritise human values, particularly the value of transparency, is crucial for promoting ethical and responsible software development practices [8]. By exploring developers' perspectives and experiences, we can gain insights into the challenges they face, and the strategies they employ to address human values violations, specifically focusing on transparency [5], [10]. Such insights can inform the development of guidelines, best practices, and educational initiatives that foster a culture of transparency in software development [9].

In this preliminary study, we aim to investigate developers' perceptions and experiences related to the human value of transparency, in the context of software application development. We explore how developers perceive transparency and its violation during the development process, and how they address and prioritise this value in their work. We also aim to contribute to the growing body of knowledge on the importance of human values in software development. By understanding the perspectives of developers and their strategies for the value of transparency, we can strive towards the creation of software applications that align with ethical standards, promote user trust, and enhance the overall societal impact of technology.

## II. BACKGROUND AND RELATED WORK

**Human Values in Software Engineering (SE):** The topic of human values in software engineering (SE) has begun to gain attention in the literature, with a focus on the ethical, social, and professional aspects of software development [1], [2], [11]. Whittle et al. make a case for considering human values such as integrity as "first-class entities" in software engineering, and calls for systematic software-engineering methods for incorporating values in the software development lifecyle [6]. However, another study reveals that software companies do consider human values in their practices, but the maturity of this consideration varies widely, depending on practitioners' awareness and organisational culture, and suggests that embedding values in technology can be achieved through an evolution of existing practices [12].

Other studies have proposed methods for measuring human values in SE. Winter et al. proposed the Values Q-sort, a systematic approach to capturing values in SE [13], while Shams et al. employed the Portrait Values Questionnaire (PVQ) to capture the values of female farmers from Bangladesh in a mobile app development project [14]. Obie et al. [15] however, argue that when designing and applying instruments for eliciting human values requirements, the specific context of the domain should be taken into account [15].

Some recent works have adopted the use of user reviews as supplementary data sources for identifying requirements related to values and their violations. Shams et al. analysed 1,522 reviews from 29 Bangladeshi agriculture apps, identifying 21 desired user values, of which 11 were reflected in the apps and 10 were missing, highlighting the importance of considering user values in app development to avoid dissatisfaction and negative socio-economic impacts [16]. Similarly, Obie et al. analysed 22,119 app reviews from the Google Play Store using natural language processing techniques, finding that 26.5% of the reviews indicated perceived violations of human values, with benevolence and self-direction being the most violated categories [17]. While [16] and [17] have focused on more general human value categories, [5] and [10] zoomed in on the specific value item of honesty - automatically detecting the violations of honesty and providing a taxonomy of the different types of honesty violations. Similar to [5] and [10], this work focuses on the single value of transparency - to understand developers' perceptions and experiences with the value of transparency, and how they approach fixing the violations of the value of transparency.

**Transparency in Software Engineering**: Transparency is an important area in software engineering (SE) and there has been some exploration of the concept of transparency in SE. Hochstetter et al. [18] introduced a transparency maturity model for government software tenders. Spagnuelo et al. argue that the transparency of a system must be considered a critical quality that must be appropriately addressed, and not simply as a high-level concept [19]. The authors proposed quality metrics for measuring transparency as a non-functional requirement for software systems. Ofem et al [8] carried out a systematic literature review on the concept of transparency in software development. Their review found that transparency remains a much under-researched non-functional quality requirement concept, especially how it might impact software development. Only three studies reviewed conceptualised transparency in software development and explored the issue of transparency as it impacts software artefacts.

Focusing on the betterment of socio-technical systems, Hosseini et al. prescribed the importance of realising transparency as a first-class requirement, as the failure of adequately implementing transparency may affect other social requirements such as privacy, trust, collaboration, and non-bias [20]. The authors further proposed a baseline model for capturing transparency requirements as an early step in this direction. Isong et al [9] propose a framework for improving the concept of transparency during software engineering. They propose a transparency improvement programme during early phases of software development along with measures of transparency in software development processes and artifacts.

Tu et al. discussed transparency within the context of SE as an attribute of communication in the development of software systems, enabling stakeholders to answer their questions about a software system during its lifecycle, and proposed accessibility, relevance, and understandability as the three key attributes for measuring transparency in SE projects [21]. The result of a survey showed that while software developers are familiar with the general concept of transparency, they are not accustomed to its practical application in software projects [21].

Other studies have focused on the social advantages of the value of transparency in SE. The results of a study with GitHub users showed that transparency in social applications in SE aids innovation, knowledge sharing, and community building [22]. Dabbish et al. argue that transparency strengthens collaboration and coordination between developers in software projects [23]. In the study of GitHub users, the authors surmise that transparency aids developers in managing their projects and deal effectively with dependencies, amongst others [23].

Another work by Tu et al. posit transparency as the visibility of information to stakeholders [7]. The results of an experiment conducted by Tu et al. show that there is a positive relationship between increased transparency of requirements and documents and more effective communication amongst various stakeholders [7].

We build on this prior body of work on transparency in SE. Our work aims to provide an overarching understanding of how developers perceive the value and violation of transparency in the software development lifecycle, and possible ways in supporting transparency in software artefacts throughout the software development lifecycle.

## III. STUDY DESIGN

### A. Aim and Research Questions

In this preliminary study, we aim to investigate developers' perceptions and experiences related to human values, with a specific focus on the value of transparency, in the context of software application development. We explore how developers perceive the value of transparency and its violations, in the development process, and how they address and prioritise these values in their work. Following this aim, we guided our study with the following three research questions:

**RQ1**: How do developers perceive the value of transparency in the development of software applications?

**RQ2**: How do developers perceive the violation of the value of transparency in the development of software applications?

**RQ3**: How do developers address reported violations of transparency?

*B. Methodology*

We followed a qualitative research methodology and conducted in-depth semi-structured interviews with 5 software practitioners to better understand their opinions on the value of transparency in software development. We present the study procedures in the following subsections. We first obtained Institutional Review Board approval for our human study (details redacted for anonymous peer review).

*1) Participant selection:* We recruited software practitioners for our study by emailing the authors' personal contacts in the industry. Participants were not compensated and participated voluntarily. In total 5 practitioners agreed to be interviewed for this preliminary study. Participants have worked in various domains and countries, with 4.8 years of professional experience on average (minimum of 2 years and maximum of 8 years). Table I summarises the demographic information of the participants.

*2) Interviewing process:* The first author conducted a series of interviews with 5 interviewees, and each interview was completed within 40 minutes. The interviews were semi-structured and divided into two parts, with more time dedicated to the latter part. In the first part, we asked some demographic questions, such as the interviewees' experience in software development, testing, and project management. In the second part, we then asked questions to understand their opinions on the human value of transparency in software development practice and showed them sample user reviews containing reports of human values violations. The interview questions were designed to explore the research questions related to the perceptions of the value of transparency, violations of transparency, and the process of fixing reported values-violations. Below are some pertinent examples of the interview questions:

  i. Do you think human values should be considered in the development of software applications? Why or why not?
  ii. What does the value of transparency mean to you as a person and as a developer?
  iii. What do you think of the violation of the value of transparency?
  iv. If several value violations are reported to you and your team, how would you prioritise which ones to fix?
  v. How would you go about fixing the violations of transparency and test that they have been fixed?

*3) Data Analysis:* We transcribed the interview recordings using Pacific Transcription Services[1] and then read the transcripts and conducted a thematic coding analysis of the transcripts. We included sentences during the coding process that are related to transparency topics. We followed the thematic analysis approach [24] to analyse and categorise the interview textual data.

The first author read the transcripts and coded the contents of the interviews using the NVIVO[2] tool for analysing the qualitative data, and discussed the codes with the second author in Zoom meetings to verify the codes and topics.

[1] https://www.pacifictranscription.com.au/
[2] https://lumivero.com/

The transcripts were interpreted in small chunks of words (codes), with recurrent codes grouped into themes. The identified themes were reviewed and refined through an iterative process. This involved examining the data within each theme, making comparisons, and ensuring that the themes accurately represented the content of the transcripts. Themes were revised as necessary to capture the variations in the data.

IV. RESULTS

In this section, we present the main themes and highlight the results of our preliminary study.

*A. RQ1: How do developers perceive the value of transparency in the development of software applications?*

*1) Transparency as a Core Value:* All participants perceive transparency as a fundamental value in software development, often closely linked with honesty and the need for accountability. They believe that being open and clear in communication and actions is essential to the process of developing software. For example, participant P1 discusses the importance of keeping stakeholders informed about progress, suggesting that transparency involves clear communication about the development process, *"...we keep in communication ensuring honesty and transparency at the same time... we would generally keep in communication to see how the project is – how the result is going on."* Participant P4 comments, *"...They [transparency] should be part of anything that we design, that's what I feel"*, while P5 says, *"...Software development, I think the first thing should be transparency I would say..."*

*2) Balancing Transparency with Practical and Ethical Considerations:* While transparency is important, it sometimes needs to be balanced with other considerations, both practical, e.g., the changing scope of a project, and ethical, e.g., respecting user privacy. Developers believe that maintaining transparency is a complex task that involves navigating various challenges and trade-offs. For example, P5 suggests that transparency is important, but sometimes certain things need to be hidden from the client due to the changing scope of the project, *"on a personal level transparency is important, but if you're really involved in the project transparency is - see, I'm not advocating to hide something, but at some stage of the project you have to hide something to the client because the scope always changes."*

*B. RQ2: How do developers perceive the violation of the value of transparency in the development of software applications?*

*1) Subjectivity of Violations:* Developers perceive a "transparency violation" as something that can vary among individuals. For example, P1 mentions that the violation of values is a grey area where the right and wrong perspectives will be different for each person interpreting the problem: *"the violation value – it is also a grey area where there is no right and wrong. So, the right and wrong perspectives will be different to each person who is interpreting the problem."* This suggests that developers recognise the complexity and subjectivity involved in identifying and addressing violations.

TABLE I
DEMOGRAPHIC INFORMATION OF THE PARTICIPANTS.

| Participant | Experience | Gender | Size of organisation | Domain | Countries of work |
|---|---|---|---|---|---|
| P1 | 2 years | Male | Small | Consulting | Australia |
| P2 | 7 years | Male | Large | Consulting | India, Australia |
| P3 | 2 years | Female | Large | Industrial Finance | Sri Lanka, Australia |
| P4 | 5 years | Male | Medium | Internet of Things | Sri Lanka, Australia |
| P5 | 8 years | Male | Large | Artificial Intelligence | India, Australia |

*2) Detecting Violations:* Developers have strategies for identifying reported transparency value violations. For example, they consider the number of similar complaints about transparency-related issues. For example, P5 suggests that if multiple people report the same complaint, it indicates that something is wrong, *"...if we get a complaint on something, if 10 people report the same complaint that means that something's wrong."* P2 corroborates this theme: *"If that particular thing is happening to a lot of users, then I'll definitely [do] a security patch saying that we encountered this problem or some users highlighted this problem here."*

*3) Consequences of Violations:* Violations of the value of transparency can have significant consequences for both developers and organisations. For example, P3 suggests that violations of transparency and honesty can harm the organisation and the individual developer, *"They might be blaming us for not giving the proper information and I believe we are not being transparent about our work... we haven't informed them. We haven't had that transparency, thereby we might be getting some cases raised from the clients."* This suggests that violations are not just theoretical issues, but can have real-world impacts. This is consistent with the results of the recent study by Obie et al. [10].

*C. RQ3: How do developers address reported violations of transparency?*

*1) Investigation and Root Cause Analysis:* Developers first engage in a process of investigation and root cause analysis. They recognise the importance of understanding the underlying factors contributing to the violation in order to effectively address it and prevent its recurrence. For example, P4 emphasises the need to investigate the root cause of a transparency violation to prevent its recurrence: *"in order to identify what's the reason behind this issue, I would do a root cause analysis to determine what's the actual issue."*

*2) Corrective Action Planning:* Developers develop corrective action plans to address reported violations of values. They believe that formulating strategies and actions are necessary to rectify the violation and prevent its future occurrence. This theme highlights the proactive planning and implementation of actions. For instance, P1 mentions the importance of developing corrective action plans to address transparency violations and ensure transparency is upheld; *"Because with transparency we let the users know what we have been up to. What kind of things we're fixing. They will know that –* *what kinds of data may have been gathered or what are the possible actions that has been done."* P3 also says, *"Each and every task within our system we had a column for criticality of that task. So we are picking the tasks by looking at the criticality of it...usually have to assign a priority or maybe criticality value for it. It would be one, two and three. So one would mean this needs to be fixed within a week. Two means it's okay if it's fixed within months. Three means it's not that much important."*

*3) Collaborative Problem-Solving:* Developers emphasise collaborative problem-solving in fixing reported values-violations. They recognise that involving relevant stakeholders, such as team members, users, or clients, leads to more effective problem-solving and resolution. Collaboration enhances the collective knowledge and expertise in addressing values violations. For example, P5 highlights the importance of considering various perspectives: *"Everyone should be treated equally because when you come to the programmers of the actual software... those are the people who will be working with the project most of the time. So the values should be valued... it's not about one particular thing on values. It is about the combination of values, then we can evaluate them and commit to a framework that we know will bring them all together in the best interests of the [project] execution."* While P3 discusses the value of involving the development team in problem-solving when addressing reported violations; *"So there's a team...They will be put into this frontline and those BAs or software engineers will be looking at these customer issues or the issues raised by the customers....all the team will get together and fix this because this is a problem that's going on their live system."*

*4) Testing and Verification:* Developers recognise the importance of testing and verification in the process of fixing reported values-violations. They believe that conducting tests and verification activities ensures that the implemented solution effectively addresses the violation and restores the desired transparency. This theme highlights the significance of ensuring the effectiveness of the implemented solution. For instance, P3 mentions the importance of testing the implemented solution to ensure that the (transparency) violation has been fixed; *"...what I feel is that when you fix that particular issue, before you push it into the production environment you could always inform quality assurance tested about this particular violation and have it as a test case."*

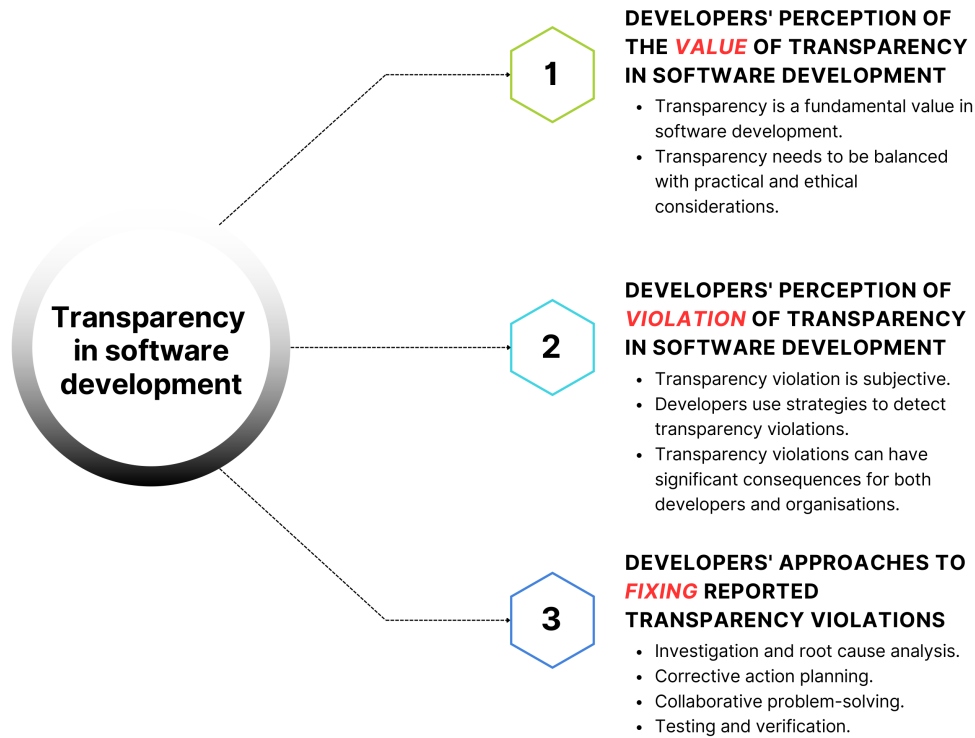

Fig. 1. The summary of the findings. How developers perceive transparency as a value in software development, how they perceive the violation of transparency, and how they address reported transparency violations.

## V. Discussion and Implications

We aimed to conduct a preliminary empirical study to explore developers' perceptions and experiences regarding human values in software development, with a specific focus on the value of transparency. Our findings shed light on several key aspects related to developers' understanding of transparency, violations of this value, and the process of addressing such reported values-violations.

Regarding RQ1, our findings revealed that developers recognise the importance of transparency as a fundamental human value in software development. They perceive transparency as crucial for building trust with users and stakeholders, promoting accountability, and fostering ethical practices. This aligns with previous research highlighting the significance of values in software development [5], [10], [16], [17].

For RQ2, we found that developers are aware of the potential violations of transparency in software applications. The themes identified were the subjectivity of values violations depending on the individuals, systemic patterns for addressing violations, and consequences of violations for both individual developers and their organisations. These findings emphasise the need for developers to proactively address and prevent such violations through ethical coding practices and robust quality assurance processes.

In answering RQ3, we uncovered several strategies employed by developers. The findings indicated that developers engage in investigation and root cause analysis to understand the underlying factors contributing to value violations. They develop corrective action plans, involve collaborative problem-solving with relevant stakeholders, and conduct testing and verification to ensure that the reported values-violations are effectively addressed. These approaches reflect the commitment of developers to rectify violations and uphold transparency in their software applications. Figure 1 summarise these findings.

Our findings have broader societal implications. Transparency in software applications is essential for building trust with users and stakeholders and ensuring the ethical and responsible use of technology. By understanding the perceptions, challenges, and strategies related to transparency, stakeholders such as regulatory bodies, policymakers, and consumer advocacy groups can develop guidelines, regulations, and standards that promote transparency in software development. This can lead to increased accountability, improved user experiences, and a more ethical and trustworthy digital environment.

## VI. Limitations

*Sample Size.* The sample size for our study is relatively small, which may limit the generalisability of the findings. Additionally, the study relied on self-reported perceptions and experiences, which are subject to biases and limitations. Future research could expand the sample size, include a more diverse range of participants, and utilise mixed-methods approaches to gain a more comprehensive understanding of the topic.

*Social Desirability Bias.* Participants may have provided responses that they believed were socially desirable, rather than fully reflecting their true perceptions and experiences.

To minimise this bias, participants were assured of the confidentiality and anonymity of their responses. The use of open-ended questions and encouraging honest and candid responses helped reduce the potential for social desirability bias.

*Researcher Bias.* The analysts' background and understanding of human values, and interpretations may have influenced the analysis and findings of this study. To address this potential bias, the analysts examined the literature on human values with a focus on the value of transparency in both the social sciences and values studies in software engineering. Furthermore, two analysts were involved in the coding and theme development process to enhance objectivity and reduce individual biases.

## VII. CONCLUSION AND FUTURE WORK

This paper explored developers' perceptions, and experiences related to human values, particularly the value of transparency, in software application development. Our findings revealed that developers highly value transparency as a fundamental human value in software development. Developers demonstrated an awareness of potential violations of transparency and acknowledged the negative impact of these violations on user trust and the overall user experience. We also provide insights into the strategies employed by developers to fix reported values-violations, including investigation and root cause analysis, corrective action planning, collaborative problem-solving, and testing and verification.

Building upon our preliminary results, there are several avenues for future research. Firstly, expanding the sample size and diversifying the participants across different software development domains, experience levels, and cultural contexts could provide a more comprehensive understanding of developers' perceptions and experiences regarding human values and transparency. Furthermore, exploring the perspectives of other stakeholders, such as users, clients, and regulatory bodies, could provide a holistic view of the significance of transparency in software development. Understanding their expectations, concerns, and experiences would contribute to the development of guidelines, standards, and policies that promote transparency and ethical practices in software application development.

## ACKNOWLEDGEMENTS

This work is supported by ARC Discovery Grant DP200100020. Madampe and Grundy are supported by ARC Laureate Fellowship FL190100035.